\documentclass[12pt,preprint]{emulateapj}

\usepackage{natbib}
\usepackage{lscape}

\shorttitle{Searching for Needles in Haystacks - Using the $Fermi$/GBM to find GRB $\gamma-$rays}

\slugcomment{Accepted for publication in the Astrophysical Journal Letters}

\begin{document}

\title{Searching for Needles in Haystacks - Using the $Fermi$/GBM to find GRB $\gamma-$rays with the $Fermi$/LAT Detector}

\author{ C.~W.~Akerlof\altaffilmark{1}, 
\email{cakerlof@umich.edu}
W.~Zheng\altaffilmark{1},
S.~B.~Pandey\altaffilmark{1,2}, 
T.~A.~McKay\altaffilmark{1}
}

\altaffiltext{1} {Randall Laboratory of Physics, Univ. of Michigan, 450 Church Street, Ann Arbor, MI, 48109-1040, USA}
\altaffiltext{2} {Aryabhatta Research Institute of Observational Sciences, Manora Peak, Nainital, India, 263129}

\shortauthors{Akerlof et al. 2010}

\begin{abstract}
From the launch of the Fermi Gamma-ray Space Telescope to July 9, 2010, the Gamma-ray Burst Monitor (GBM) has detected 497 probable GRB events. Twenty-two of these satisfy the simultaneous requirements of an estimated burst direction within $52^\circ$ of the Fermi Large Area Telescope (LAT) boresight and a low energy fluence exceeding 5 $\mu$erg/cm$^2$. Using matched filter techniques, the spatially correlated $Fermi$/LAT photon data above 100 MeV have been examined for evidence of bursts that have so far evaded detection at these energies. High energy emission is detected with great confidence for one event, GRB 090228A. Since the LAT has significantly better angular resolution than the GBM, real-time application of these methods could open the door to optical identification and richer characterization of a larger fraction of the relatively rare GRBs that include high energy emission.
\end{abstract}

\keywords{Gamma-ray burst: general}

\section{Introduction}

One of the more surprising results of the Compton Gamma-Ray Observatory (CGRO) was the EGRET discovery of an 18 GeV photon associated with GRB 940217 (Hurley 1994). About a half-dozen bursts were seen over the course of the CGRO mission with photons above 100 MeV (Catelli 1998, Dingus 2003). Since the GRB spectral energy distribution at lower energies has been well characterized by a modified power law with peak fluxes at energies of the order of 200 KeV, the existence of photons at energies $10^4$ times higher puts a significant constraint on any viable model of the GRB phenomenon. This has been
a subject of great interest for missions that followed EGRET. Prior to launch of the Fermi Gamma-ray Space Telescope, it was possible to speculate that the LAT instrument would detect more than 200 GRB events per year (Dingus, 2003). In the two year period since the launch of the Fermi Gamma-ray Space Telescope, the Gamma-ray Burst Monitor (GBM) has reported approximately 475 GRBs, ie. a rate of about 250 per year. Over essentially the same period, only 17 bursts have been identified by the Fermi/LAT. We now see that the range of GRB photon energies extends over a scale of $10^6$ but
the physical dynamics of these phenomena are still not understood. This is coupled to the question of whether high energy photons are associated with all GRBs or only with a small sub-class. Since the Fermi mission is unlikely to be duplicated any time soon, there is some urgency to assuring that the maximum information is being extracted from
this valuable facility. Thus, our group has set about
developing techniques for enlarging the number of gamma-ray bursts identified with high energy photon emission, ie. above 100 MeV.

The first result of this effort has established the correlation of two $Swift$/XRT-localized bursts, GRB 080905A and
GRB 091208B, with high energy photons in the $Fermi$/LAT detector (Akerlof el al. (2010), hereafter A10). The statistical technique
employed is the $matched~filter$ method, most familiar to those detecting signals in the time domain. The underlying assumption is that the characteristics of both the signal and background are $a~priori$ known functions of one or more variables. Since the matched filter maximizes the signal-to-noise ratio, moderate departures from optimality degrade the filter performance relatively slowly, making this a valuable tool for investigating the possible existence of faint signals. The details of the filter algorithm are explicitly described in A10. In this paper, we take the next harder step of dropping our reliance on precision burst coordinates provided by
Swift or other similar high resolution instruments. Instead, we use the approximate localization of the $Fermi$/GBM
to map a region of interest on the $Fermi$/LAT field of view. By identifying high energy photon clusters, provisional
burst coordinates can be determined with significantly smaller errors than available from the GBM. From there, the
burst identification follows along lines set out in A10.

\section{Sample Selection}

As a first step in this program, a list of all GBM triggers was obtained from the
$fermigbrst$ catalog maintained by the $Fermi$ Science Support
Center\footnote{http://heasarc.gsfc.nasa.gov/W3Browse/fermi/fermigbrst.html}. The
catalog contains 497 GRB triggers from launch to July 9, 2010.  This list was
cross-matched with Table 1 in Guetta \& Pian (2009) and Table 2 in Guetta et al.
(2010) to identify the burst GCN designations and the low energy fluences.
For triggers occuring after February 18, 2010, fluences were obtained from individual GCN circulars.
GBM triggers were also checked against XRT locations from
$Swift$\footnote{http://heasarc.gsfc.nasa.gov/docs/swift/archive/grb$\_$table/} to remove
events already considered in A10.

Using data from the Fermi spacecraft attitude file, we further selected those triggers with a boresight angle
less than 52$^\circ$ and an estimated GBM error circle less than 10$^\circ$. Events without
GBM fluence information or previously claimed LAT detections\footnote{
http://fermi.gsfc.nasa.gov/ssc/observations/types/grbs/grb$\_$table/} were also discarded.
Applying a final cut on GBM fluence (8 - 1000 KeV) requiring greater than 5.0 $\mu$erg/cm$^2$ reduced
the number to 22 events (see Table 1). These are termed the ``GBM'' data. 464 additional fields were
taken at random on the sky
with similar criteria to study the background behavior and are identified as the``random'' data in the following text.

\begin{deluxetable}{llrrr}
 \tabcolsep 0.4mm
 \tablewidth{0pt}
 \tablecaption{List of 22 GBM trigger GRBs}
  \tablehead{\colhead{GRB} & \colhead{Trigger} & \colhead{RA} & \colhead{Dec} & \colhead{$S_{GBM}$ } \\
  \colhead{} & \colhead{} & \colhead{($^{\circ}$)} & \colhead{($^{\circ}$)} & \colhead{$\mu$erg/cm$^2$} }
\startdata
080830  & 080830368  & 160.10  &  30.80  &   9.2 \\
080904  & 080904886  & 214.20  & -30.30  &   5.0 \\
080906B & 080906212  & 182.80  &  -6.40  &  10.9 \\
080925  & 080925775  &  96.10  &  18.20  &  19.4 \\
081122A & 081122520  & 339.10  &  40.00  &   9.6 \\
081231  & 081231140  & 208.60  & -35.80  &  12.0 \\
090112A & 090112332  & 110.90  & -30.40  &   5.2 \\
090131  & 090131090  & 352.30  &  21.20  &  22.3 \\
090227A & 090227310  &   3.30  & -43.00  &   9.0 \\
090228A & 090228204  & 106.80  & -24.30  &   6.1 \\
090319  & 090319622  & 283.30  &  -8.90  &   7.5 \\
090330  & 090330279  & 160.20  &  -8.20  &  11.4 \\
090514  & 090514006  &  12.30  & -10.90  &   8.1 \\
090516B & 090516137  & 122.20  & -71.62  &  30.0 \\
090829A & 090829672  & 329.23  & -34.19  & 102.0 \\
090829B & 090829702  & 354.99  &  -9.36  &   6.4 \\
090922A & 090922539  &  17.16  &  74.30  &  11.4 \\
091120  & 091120191  & 226.81  & -21.79  &  30.2 \\
100122A & 100122616  &  79.20  &  -2.71  &  10.0 \\
100131A & 100131730  & 120.40  &  16.45  &   7.7 \\
100423B & 100423244  & 119.67  &   5.78  &  12.3 \\
100511A & 100511035  & 109.29  &  -4.65  &   7.1 \\
\enddata
\end{deluxetable}

\section{Signal Detection Technique}

The core task of this search procedure is the identification of triplet clusters of photons in the $Fermi$/LAT instrument
whose spatial accuracy is considerably better than the GBM. The set of photon data for each candidate
burst is confined to lie within a $16^\circ$ cone angle of the GBM direction and a time window extending from zero to
47.5 s after the GBM burst trigger. The procedure first computes a signal weight for each photon pair based on
photon energy, detection time, photon event class and angular separation relative to the expected LAT PSF errors. The
photon pair weights are subject to a weak threshold cut designed to avoid combinatorial overload should large
photon numbers be encountered. In practice, this was not a severe problem and can be ignored. The formula for the pair
weights is given by:
\begin{equation}
Q_{ij} = w_i{\ }{\cdot}{\ }w_j{\ }{\cdot}{\ }\Delta_{ij},
\end{equation}
where
\begin{equation}
w_i = w_E(i){\ }{\cdot}{\ }w_t(i){\ }{\cdot}{\ }w_c(i){\ }{\cdot}{\ }4\pi {\sigma}^2_{PSF}(E_i),
\end{equation}
\begin{equation}
\Delta_{ij}=\frac{e^{-\delta_{ij}}}{4\pi ({\sigma}^2_{PSF}(E_i) + {\sigma}^2_{PSF}(E_j))},
\end{equation}
\begin{equation}
\delta_{ij} = \frac{1}{2}\frac{{\theta}^2_{ij}}{{\sigma}^2_{PSF}(E_i) + {\sigma}^2_{PSF}(E_j)},
\end{equation}
and ${\theta}_{ij}$ is the angle between the $i$'th and $j$'th photon. The definitions of $w_E$, $w_t$
and $w_c$ can be found in equation 1,3 and 4 of A10.

The next step is to link photon pairs so that the three pairs, \{i, j\}, \{j, k\} and \{i, k\}, become identified
as the triplet, \{i, j, k\}. The triplet weight value is computed by the formula:
\begin{equation}
R_{ijk} = (w_i{\ }{\cdot}{\ }w_j{\ }{\cdot}{\ }w_k{\ }{\cdot}{\ }\Delta_{ij}{\ }{\cdot}{\ }\Delta_{jk}{\ }{\cdot}{\ }\Delta_{ik})^{\frac{1}{3}}
\end{equation}

The triplet weights are ranked by value and the set is pruned by the condition that a triplet element, $R_{ilm}$, is removed if $R_{ilm} < R_{ijk}$. This leaves a set of triplet clusters, each with a discrete complement of three photons. For
each of these clusters, a PSF-weighted estimate of the burst direction is performed and the matched filter angle weight,
$w_\theta$ is computed with respect to this vector as described in equation 2 of A10. At this point, an event weight
for each cluster is computed by the formulas given in A10 with one small modification. In the scheme described here,
the GRB direction is not initially defined with any precision. Thus, it is inappropriate to include a $1/\sigma^2_{PSF}$
factor for all values of $w_E$. For the highest energy photon in each triplet, the $4\pi\sigma_{PSF}^2$ factor is
removed to reflect that this leading photon plays the principal role in fixing the apparent GRB direction. Although the
calculations carry each cluster through the same computational path, the expectation is that the cluster with the highest
matched filter weight is the most probable identification.

The 22 GBM fields described earlier were the target of this investigation. We recognized
that the most convincing argument for a true LAT identification should rely on the statistical distributions
for the matched filter weights in LAT fields with similar characteristics. To increase that number
as much as possible, the LAT fields of view were segmented into 12 circular tiles embedded on a spherical surface. Each tile subtends a cone with a half-angle of 16.0$^\circ$. This tiling scheme was applied
to both the GBM and random field data sets to realize 182 and 3440 independent directions in space satisfying all the
criteria described previously. Taking advantage of the fact that each field observation was blocked into a 250-s segment,
the number of independent observations was multiplied by five by regarding each 50-s time slice as a separate sample.
Thus, there are 910 background measurements taken from LAT observations obtained simultaneously with the
candidate GBM fields and an additional 17200 samples taken under similar but not identical conditions. One particular
concern for an analysis of this type is that false positives will selectively occur as the sample photon rate
rises substantially above the mean. Evidence that this is not the case here is shown in Figure 1 which plots the
cumulative distribution of the total number of photons within the LAT field of view over a 250-s interval. These rates
explicitly exclude contamination from photons beyond the 105$^\circ$ zenith angle cut. As shown in the plot,
the prominent GBM trigger event reported here is not associated specifically with fields with high ambient
background rates. The
similarity of the distributions for GBM and random fields also shows that the GBM data are not pathological as far as
rates are concerned.


\begin{figure}
  \includegraphics[scale=0.475]{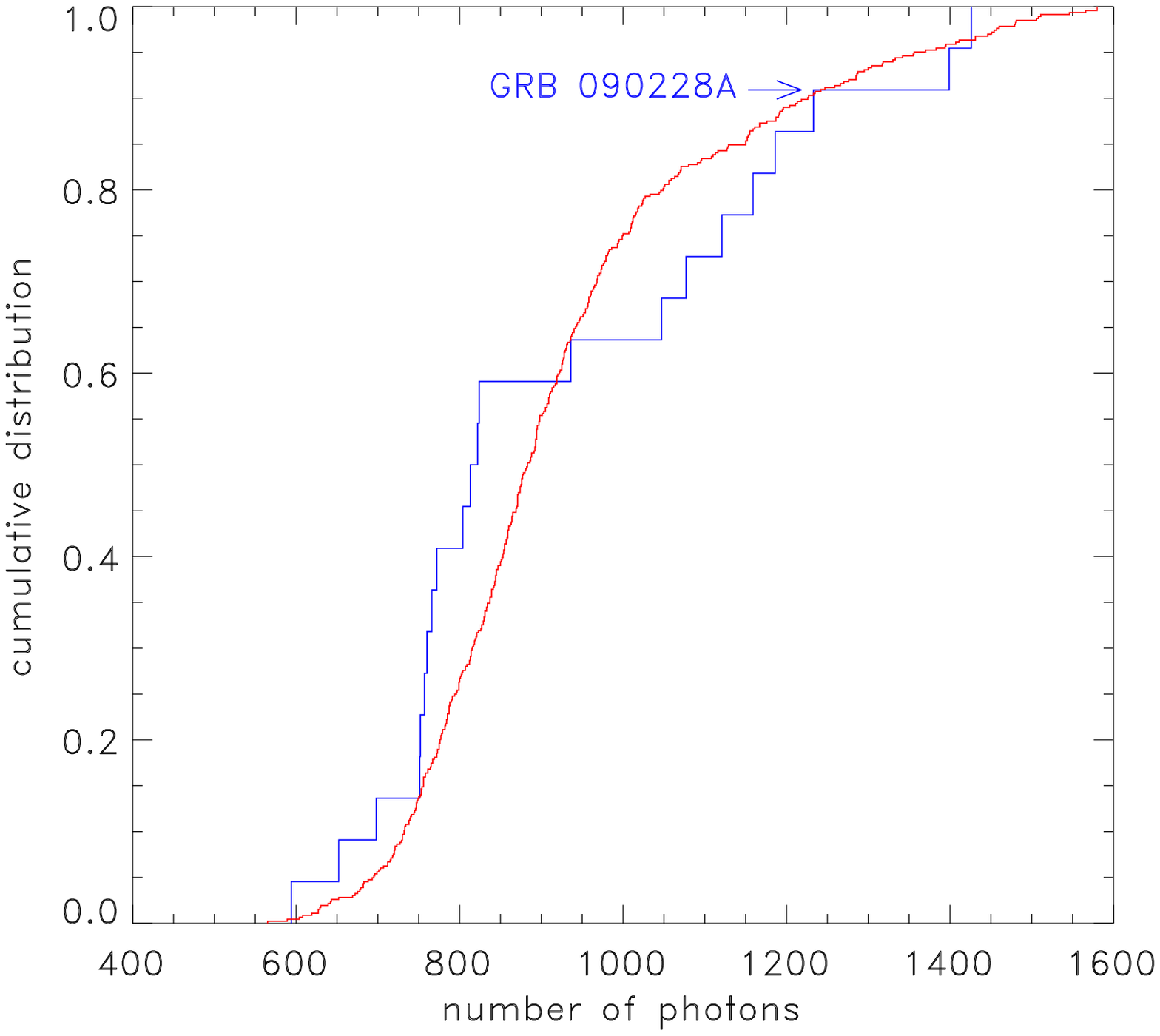}
  \caption{Cumulative distributions of LAT photon rates over 250-s intervals for the GBM (blue) and random
background (red) fields. These rates reflect the entire LAT FoV except for photons that lie outside the
105$^\circ$ zenith angle cut. The rate corresponding to the most prominent GBM event is indicated by the arrow.}
\end{figure}

\section{Results}

Our statistical localization and weighting scheme identified one outstanding candidate for high
energy photon emission, GRB 090228A. The best estimate for the
probability of such an occurence by chance alone was obtained by performing identical searches on
random LAT fields with the same criteria. Thus, 11 out of 17200 random fields generated matched filter weights
exceeding the value for our candidate event. Multiplying by a trials factor of 22 for the number of GBM localized fields considered yields a false positive probability of 1.4\%. To check that these correlations were simply not due
to preferentially higher background rates for the GBM exposures, we also performed similar calculations for
the LAT data confined to an average of 8 uncorrelated directions per exposure and five independent time intervals from the same GBM data sets.
In this case, 2 fields out of 910 exceed the candidate signal for a false positive rate of 4.8\%. The cumulative distributions are plotted in Figure 2. The statistical similarity of the
GBM off-axis and random field data demonstrates that the GBM data set is not correlated with anomalous environmental
conditions such as higher cosmic ray background rates. A list of photons associated with this burst is provided in Table 2.

The GBM data for GRB 090228A is described in GCN 8918 (von Kienlin, et al., 2009). According to this note, the
burst was localized to a $1-\sigma$ accuracy of better than $1^\circ$ with an additional systematic uncertainty
of the order of $2.5^\circ$. The coordinate values obtained by the GBM group and this analysis are listed in Table 2.
The GBM value generated some concern since our cluster finder position disagreed by 8.7$^\circ$. Fortuitously
as this manuscript was being drafted, a paper was posted to astro-ph (Guiriec et al., 2010) providing a burst direction
with an estimated accuracy of $0.2^\circ$ and lying $0.5^\circ$ from our own estimate. We believe that this establishes
the validity of our identification to near certainty. In addition, the positive GRB correlation of Event Class 2 and 3
photon rates discussed in A10 was observed for the ensemble of GBM fields as well. The photon clustering is easily observed in the sky map shown in Figure 3.

\begin{figure}
  \includegraphics[scale=0.475]{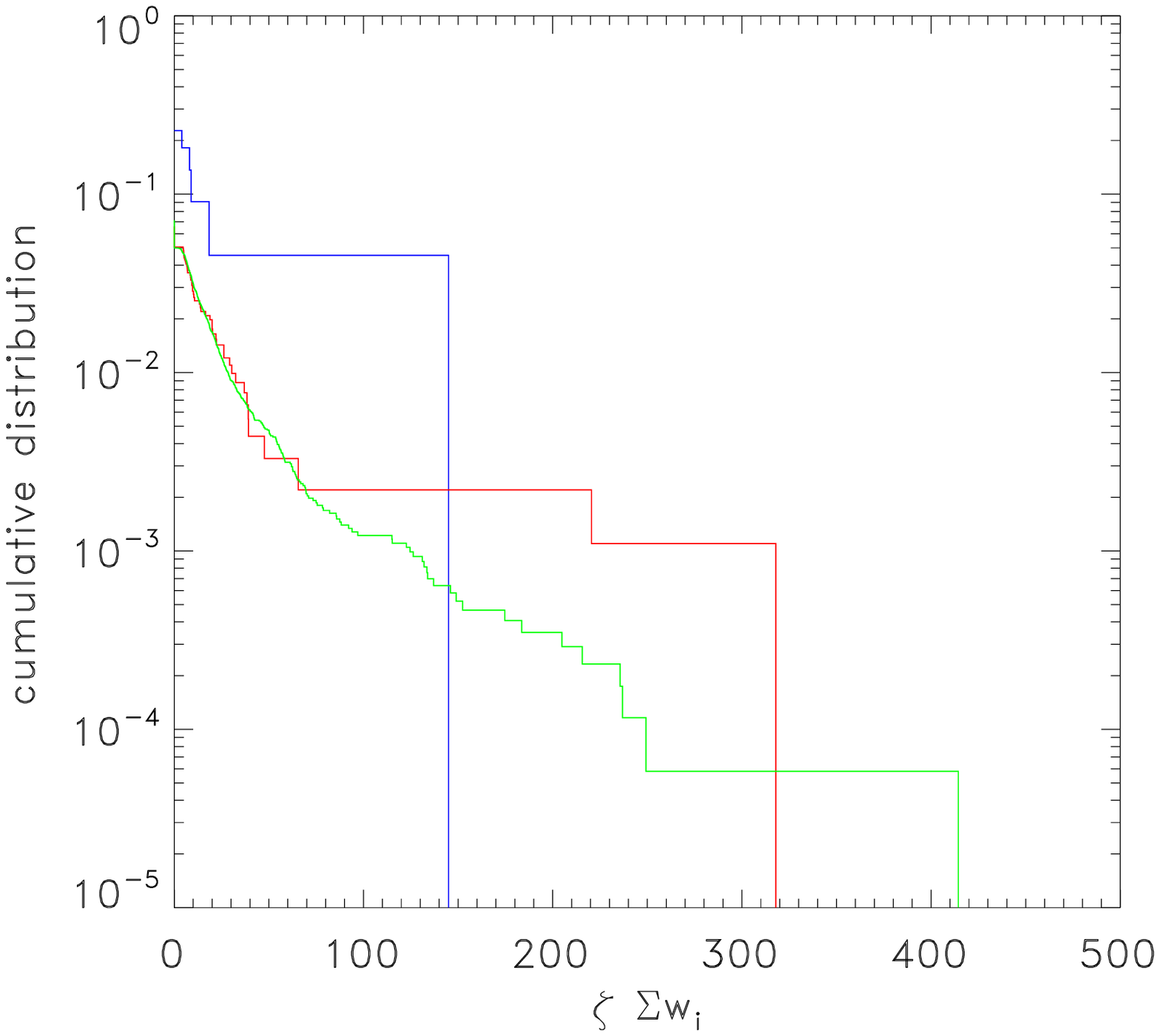}
  \caption{Complements of the cumulative distributions for ${\zeta}{\sum}w_i$
           for 22 GBM fields (blue), 910 random fields obtained nearly simultaneously with
           the GBM data (red) and 17200 random fields obtained at random times (green).}
\end{figure}

\begin{deluxetable}{crrrrr}
 \tabcolsep 0.4mm
 \tablewidth{0pt}
 \tablecaption{~ GRB 090228A high energy photon list and celestial coordinate estimates}
  \tablehead{\colhead{$i$} & \colhead{$t$} & \colhead{$\theta$} & \colhead{$E$} & \colhead{$c$} & \colhead{$w_i$} \\
  \colhead{} & \colhead{(s)} & \colhead{($^{\circ}$)} & \colhead{(MeV)} & \colhead{} & \colhead{} }
\startdata
  1  &     2.007   &    1.692  &   125.241 & 2  &    52.814   \\
  2  &     3.752   &    2.611  &   206.983 & 3  &    46.545   \\
  3  &    25.141   &    0.592  &   308.638 & 3  &    44.278   \\
  4  &     3.243   &    1.978  &   638.692 & 3  &     1.356   \\
  5  &     4.966   &    0.063  &  2787.028 & 3  & $^*$0.376   \\
  6  &    33.621   &    5.063  &   340.623 & 1  &     0.002   \\
\\
\multicolumn{6}{c}{$\zeta$ = 0.99722~~~~~$\zeta\sum w_i$ = 144.969} \\
\\
\hline
\hline
\\

& source & $\alpha$   & $\delta$   & $\sigma_\theta$ & $\theta_{i-1, i}^d$  \\
&        & ($^\circ$) & ($^\circ$) & ($^\circ$)      &   ($^\circ$)         \\
\\
\hline
\\
& GBM$^a$  &     106.80   &    -24.30  &   $\lesssim3.0$   &   \\
& LAT$^b$  &      98.56   &    -28.86  &   0.18         & 8.66 \\
& IPN$^c$  &      98.30   &    -28.40  &   0.02         & 0.51 \\
\enddata
\tablenotetext{*}{indicates diminished $w_E$ for highest energy photon}
\tablenotetext{a}{von Kienlin et al. (2009)}
\tablenotetext{b}{this paper}
\tablenotetext{c}{Guiriec et al. (2010)}
\tablenotetext{d}{$\theta_{i-1, i}$ is the angle between the spatial directions for the GBM and LAT or
the LAT and IPN}
\end{deluxetable}

\begin{figure}
  \includegraphics[scale=0.475]{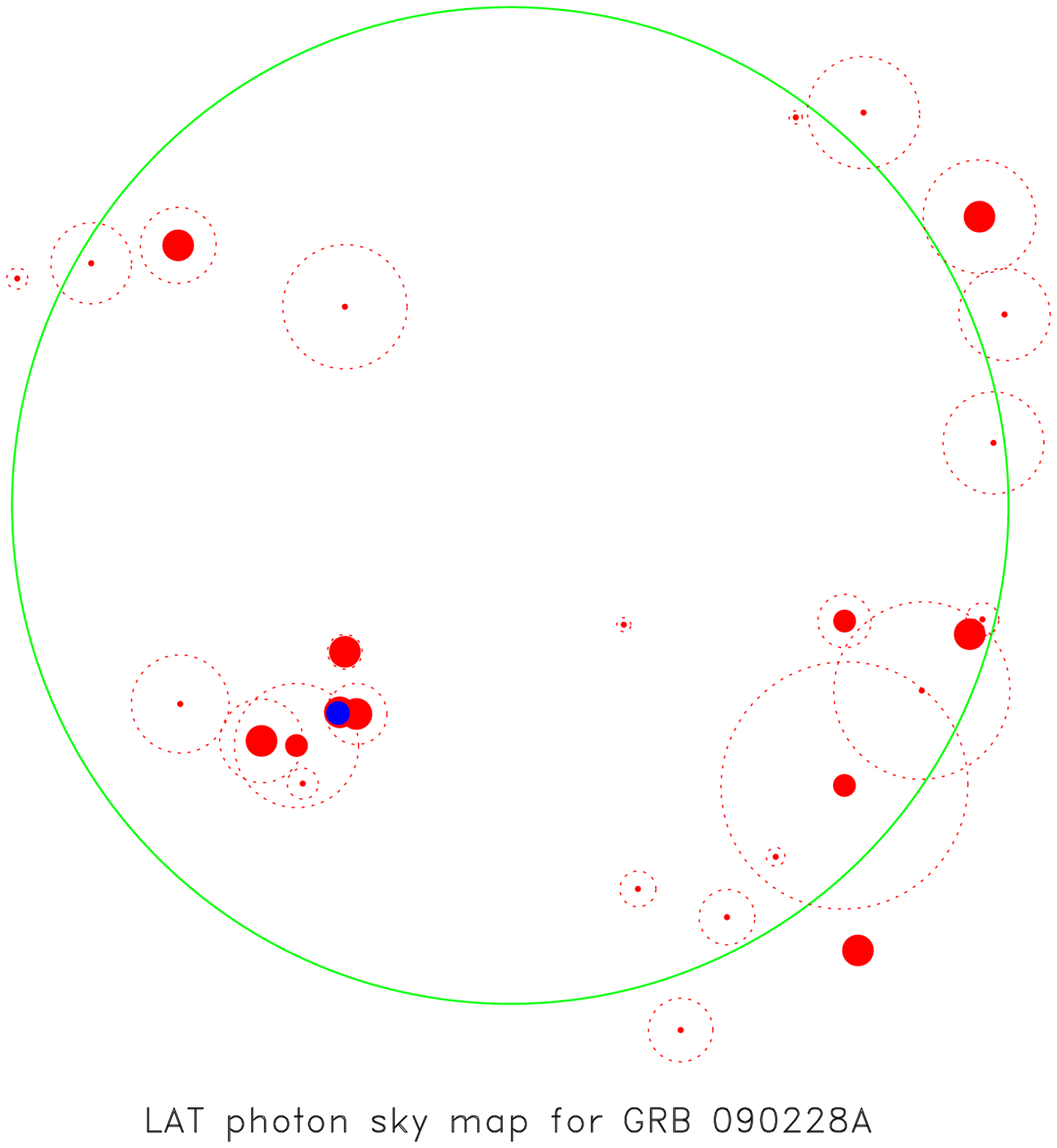}
  \caption{Sky map of $>100$ MeV photons for GRB 090228A. The diameter of each dot is proportional to its
statistical weight. Thus, the largest diameters represent Event Class 3, etc. The dotted circles
around each point indicate the $1-\sigma$ errors. The figure is centered on the nominal
coordinates furnished by the GBM; the blue dot on the lower left shows the GRB coordinates computed by the
cluster algorithm described in the text. The large green circle depicts the boundaries of the 16.0$^\circ$ cone
that defines the fiducial boundaries for the cluster search. North is up and East is to the right.}
\end{figure}

\section{Discussion}

The one event identified in this paper establishes the validity of our statistical techniques to a level of near certainty. By using GBM triggers to guide the discovery of photon clusters in the LAT, the phase space
for finding counterparts can be reduced from hundreds of square degrees to one square degree or less. This
makes a very significant difference for those seeking to identify GRB optical counterparts. If the algorithms
used here could be adapted to the real time environment, the number of bursts with high energy associations
could be increased appreciably. The additional computational load is negligible - about 30 ms per day.
For such real time applications, the high selectivity employed here is overkill - any
identification that can be corroborated optically will suffice. Thus, effective signal-to-noise rates of the order of unity are extremely valuable. As shown here, these techniques greatly enhance the dynamic range over which high
energy radiation can be explored. It is not too outrageous to claim that this is the equivalent of making the LAT
three to ten times larger in size.

As was noted in A10, the most surprising aspect of our recent work is the very small number of GRBs that can be positively identified with high energy emission despite the substantially lower fluence thresholds. This is a mystery which deserves serious consideration. We hope that raising new questions is sometimes more useful than answering old
problems.

\vspace{0.1cm}

\acknowledgments
 
We thank Chris Shrader, director of the $Fermi$ Science Support Center for his considerable help in obtaining and interpreting the $Fermi$ mission data products. Fang Yuan provided valuable assistance in the initial process of learning
how to access and manipulate the $Fermi$ data. This research is supported by NASA grant NNX08AV63G and NSF
grant PHY-0801007.

\end{document}